\documentclass[english,10pt,aps,prd,a4paper,preprintnumbers,floatfix,nofootinbib,showpacs,superscriptaddress, notitlepage,twocolumn]{revtex4-1}
\pdfoutput=1
\usepackage[english]{babel}
\usepackage{amsmath}
\usepackage{graphicx}
\usepackage{color}
\usepackage{mathrsfs}   
\usepackage{amssymb}
\usepackage{hyperref}
\usepackage{enumerate}
\usepackage[normalem]{ulem} 
\usepackage{dcolumn}
\usepackage{bm}
\usepackage{graphicx}
\usepackage[sort&compress]{natbib}
\usepackage{epstopdf}
\usepackage{subcaption}
\usepackage{multirow}
\usepackage{rotating}
\usepackage[utf8]{inputenc} 
\usepackage{enumitem,xcolor}

\newcommand {\be} {\begin{equation}}
\newcommand {\ee} {\end{equation}}

\definecolor{greenLinks}{rgb}{0, 0.6, 0} 
\definecolor{blueLinks}{rgb}{0, 0, 0.6}
\definecolor{redLinks}{rgb}{0.6, 0, 0}
\definecolor{tempText}{rgb}{0.55, 0.10,0.67}
\definecolor{eprintLinks}{rgb}{0.4, 0.4, 0.4}
\definecolor{journalLinks}{rgb}{0.6, 0, 0}

\def\vev#1{\left\langle #1\right\rangle}

\usepackage{float}

\def\vev#1{\left\langle #1\right\rangle}

\def\21{$\mathrm{SU(2)_L \otimes U(1)_Y}$ }
\def\31{$\mathrm{SU(3)_c \otimes U(1)_Q}$ }

\def\3211{$\mathrm{SU(3) \otimes SU(2)_L \otimes U(1)_R \otimes U(1)_{B-L}}$ }
\def\321{$\mathrm{SU(3) \otimes SU(2) \otimes U(1)}$ }
\def\422{$\mathrm{SU(4) \otimes SU(2) \otimes SU(2)_R}$ }
\newcommand {\ignore}[1]{}

\def\vev#1{\left\langle #1\right\rangle}

\newcommand{\AddrAHEP}{%
  AHEP Group, Institut de F\'{i}sica Corpuscular --
  CSIC-Universitat de Val\`{e}ncia, Parc Cient\'ific de Paterna.\\
 C/ Catedr\'atico Jos\'e Beltr\'an, 2 E-46980 Paterna (Valencia) - SPAIN}
\newcommand{\AddrUNAM}{ {\it Instituto de F\'{\i}sica, Universidad Nacional Aut\'onoma de M\'exico, A.P. 20-364, Ciudad de M\'exico 01000, M\'exico.}}
\newcommand{\AddrTUM}{ Physik-Department T30d, Technische Universit\"{a}t M\"{u}nchen.\\
James-Franck-Strasse, 85748 Garching, Germany}
\newcommand{\AddrChile}{Departamento de F\'isica, Universidad Cat\'olica del Norte, Avenida Angamos 0610, Casilla 1280, Antofagasta, Chile.}
\newcommand{\AddrIndia}{Department of Physics, Indian Institute of Science Education and Research, Bhopal Bypass Road, Bhauri, Bhopal, India.}

\begin{document}

\title{Dark matter stability and Dirac  neutrinos using only Standard Model symmetries}
\author{Cesar Bonilla}\email{cesar.bonilla@tum.de}
\affiliation{\AddrTUM}
\affiliation{\AddrChile}
\author{Salvador Centelles-Chuli\'{a}}\email{salcen@ific.uv.es}
\affiliation{\AddrAHEP}
\author{Ricardo Cepedello}\email{ricepe@ific.uv.es}
\affiliation{\AddrAHEP}
\author{Eduardo Peinado}\email{epeinado@fisica.unam.mx}
\affiliation{\AddrUNAM}
\author{Rahul Srivastava}\email{rahul@iiserb.ac.in}
\affiliation{\AddrAHEP}
\affiliation{\AddrIndia}

\begin{abstract}
  \vspace{1cm} 
We provide a generic framework to obtain stable dark matter along with naturally small Dirac neutrino masses generated at the loop level. This is achieved through the spontaneous breaking of the global $U(1)_{B-L}$ symmetry  already present in the Standard Model. 
The $U(1)_{B-L}$ symmetry is broken down to a residual even $\mathcal{Z}_n$ ($n \geq 4$) subgroup. The residual  $\mathcal{Z}_n$ symmetry simultaneously guarantees dark matter stability and protects the Dirac nature of neutrinos. 
The $U(1)_{B-L}$ symmetry in our setup is anomaly free and can also be gauged in a straightforward way.
Finally, we present an explicit example using our framework to show the idea in action.

\end{abstract}

\maketitle

At present, a plethora of cosmic observations all indicate that the bulk of matter in the Universe is in the form of dark matter, a hitherto unknown form of matter which interacts gravitationally, but has little or no electromagnetic interaction \cite{Aghanim:2018eyx}. 
Similarly, the observation of neutrino oscillations has conclusively proven the existence of mass for at least two active neutrinos \cite{deSalas:2017kay, Whitehead:2016xud, Abe:2017uxa, Decowski:2016axc}. 
These observations are two of the most serious shortcomings of the Standard Model (SM) since in the SM there is no viable candidate for dark matter and neutrinos are predicted to be massless.
Thus, they both inarguably point to the presence of new physics beyond the SM and they are topics of active theoretical and experimental research.

To explain dark matter, the particle content of the SM needs to be extended. Furthermore, to account for dark matter stability new explicit \cite{ Silveira:1985rk, Ma:2006km} 
or accidental symmetries \cite{Cirelli:2005uq} beyond those of the SM are also invoked.
On the other hand, the understanding of the tiny, yet non-zero, masses of neutrinos also requires extending the SM in one way or another 
\cite{Ma:1998dn, CentellesChulia:2018gwr}. 
However, the type of SM extensions required to explain the neutrino masses depend crucially on the Dirac/Majorana nature of neutrinos.
This still remains an open question despite a tremendous amount of experimental effort \cite{GERDA:2018zzh,KamLAND-Zen:2016pfg,Agostini:2017iyd,Kaulard:1998rb}. 

There are several ongoing and planned experiments searching for neutrinoless double beta decay \cite{KamLAND-Zen:2016pfg,Agostini:2017iyd}, which if observed, owing to the Black-Box theorem~\cite{Schechter:1981bd}, would imply that neutrinos are Majorana particles. 
Furthermore, inference about nature of neutrinos can also be derived if lepton number violating decays are observed at colliders \cite{Aaboud:2018spl} or in the conversion of $\mu^- \to e^+$ in muonic atoms \cite{Kaulard:1998rb}
In the absence of any experimental or observational signature, the nature of neutrinos remains an open question.

From a theoretical point of view, the issue of the Dirac/Majorana nature of neutrinos is intimately connected with the $U(1)_{B-L}$ symmetry of the SM and its possible breaking pattern \cite{Hirsch:2017col}.
If the $U(1)_{B-L}$ symmetry is conserved in nature, then the neutrinos will be Dirac fermions. 
However, if it is broken to a residual $\mathcal{Z}_m$ subgroup with $m\in \mathbb{Z}^+$ and $m \geq 2$, with $\mathbb{Z}^+$ being the set of all positive integers, then the Dirac/Majorana nature will depend on the residual $\mathcal{Z}_m$ symmetry provided that the SM lepton doublets $L_i = (\nu_{L_i}, l_{L_i})^T$  do not transform trivially under it. Thus, we have
\begin{eqnarray}
U(1)_{B-L}   & \, \to  \, &   \mathcal{Z}_m \equiv \mathcal{Z}_{2n+1} \, \text{with} \,  n \in \mathbb{Z}^+    \nonumber \\
& \, \Rightarrow \, & \text{neutrinos are Dirac particles}      \nonumber \\
U(1)_{B-L}  & \, \to  \,  & \mathcal{Z}_m \equiv \mathcal{Z}_{2n} \, \text{with} \,  n \in \mathbb{Z}^+  \\
& \, \Rightarrow \, & \text{neutrinos can be Dirac or Majorana } \nonumber
\label{oddzn}
\end{eqnarray}
If the $U(1)_{B-L}$ is broken to a $\mathcal{Z}_{2n}$ subgroup, then one can make a further classification depending on how the $L_i$ transform,
\begin{eqnarray}
 L_i \left\{ \begin{array}{ll}
          \nsim  \omega^{n} \ \ \text{under $\mathcal{Z}_{2n}$} &  \Rightarrow\text{Dirac neutrinos}\\
          \sim  \omega^{n}\ \ \text{under $\mathcal{Z}_{2n}$} &  \Rightarrow\text{Majorana neutrinos}\end{array} \right. 
\label{evenzndir}
\end{eqnarray}
where $\omega^{2n}=1$. 
Thus, from a symmetry point of view there are more options that lead to Dirac neutrinos than Majorana.
Moreover, in some recent works it has been argued that in a full theory with the weak gravity conjecture neutrinos are expected to be Dirac fermions \cite{Ibanez:2017kvh}.
Owing to the above arguments, in the past few years, Dirac neutrinos have gained a significant amount of attention leading to the development of several elegant mass mechanisms to generate naturally small Dirac neutrino masses \cite{Ma:2014qra,Ma:2015mjd,Chulia:2016giq,Bonilla:2016diq,Ma:2016mwh,Wang:2016lve,Borah:2017dmk,Yao:2018ekp,Reig:2018mdk,Han:2018zcn}.

Coming back to dark matter, there are particularly attractive scenarios that connect dark matter to neutrino physics in an intimate manner. The scotogenic model is one such model where the ``dark sector'' participates in the loop responsible for neutrino mass generation \cite{Ma:2006km}. Recently, a relation between the Dirac nature of neutrinos and dark matter stability has also been established \cite{Chulia:2016ngi}. 
Furthermore, it has been shown that this relation is independent of the neutrino mass generation mechanism \cite{CentellesChulia:2018gwr}. 
It utilizes the SM lepton number $U(1)_L$ symmetry\footnote{One can equivalently use the anomaly free $U(1)_{B-L}$ symmetry.}, or its appropriate $\mathcal{Z}_n$ subgroup, to forbid Majorana mass terms of neutrinos as well as to stabilize dark matter \cite{Chulia:2016ngi}.
In this approach, the Dirac nature of neutrinos and the stability of dark matter are intimately connected, having their origins in the same lepton number symmetry.

In this paper we aim to combine and generalize these two approaches and develop a general formalism where the following conditions are satisfied:

\begin{enumerate}
\item[ I.] Neutrinos are Dirac in nature.
\item[ II.]   Naturally small neutrino masses are generated through finite loops, forbidding the tree-level neutrino Yukawa couplings.
\item[ III.] The dark sector participates in the loop. The lightest particle being stable is a good dark matter candidate.\\
\end{enumerate}

Usually one needs at least three different symmetries besides those within the Standard Model to achieve this \cite{Bonilla:2016diq}.
However, we show that all of these requirements can be satisfied without adding any extra explicit or accidental symmetries. In our formalism we employ an anomaly free chiral realization of the  $U(1)_{B-L}$ spontaneously broken to a residual $\mathcal{Z}_n$ symmetry and show that just the $U(1)_{B-L}$ already present in the SM is sufficient.

Before going into the details of the formalism, let us briefly discuss the possibility of chiral solutions to $U(1)_{B-L}$ anomaly cancellation conditions.
It is well known that the accidental $U(1)_B$ and $U(1)_L$ symmetries of the SM are anomalous, but the $U(1)_{B-L}$ combination can be made anomaly free by adding three right-handed neutrinos $\nu_{R_i}$ with $(-1,-1,-1)$ vector charges under $U(1)_{B-L}$. 
However, chiral solutions to $U(1)_{B-L}$ anomaly cancellation conditions are also possible. The particularly attractive feature of chiral solutions is that by using them one can automatically satisfy conditions I and II, as shown in \cite{Ma:2014qra,Ma:2015mjd}, using the chiral solution $\nu_{R_i} \sim (-4,-4,5)$ under $U(1)_{B-L}$ symmetry.

\begin{widetext}

 \begin{figure}[th]
    \centering
    \begin{subfigure}[b]{0.43\textwidth}
        \includegraphics[width=\textwidth]{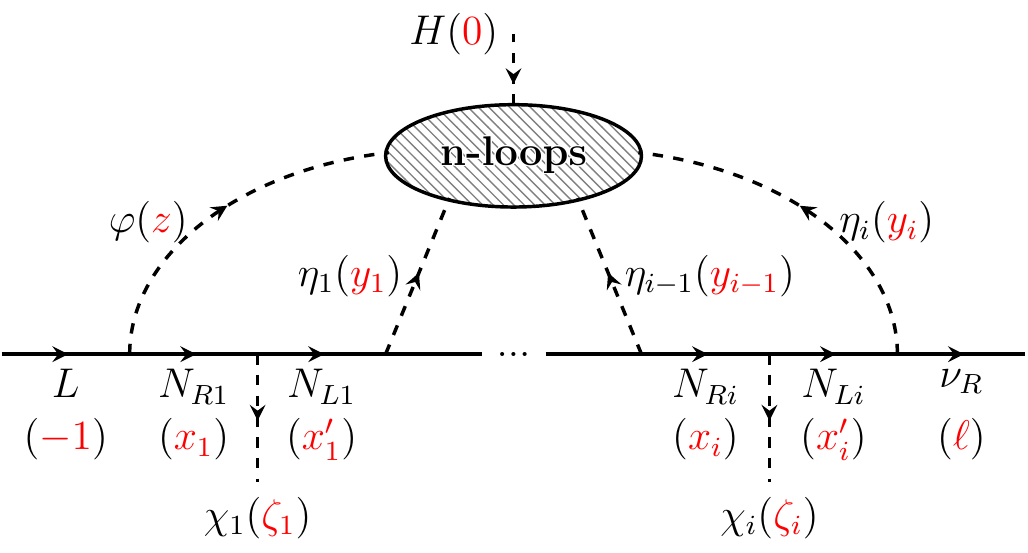}
        \caption{General $U(1)_{B-L}$ charge assignment.}
        \label{fig:genU1}
    \end{subfigure}
    ~ 
    \begin{subfigure}[t]{0.1\textwidth}
    \vspace*{-4.15cm}
    \hspace*{-0.45cm}
   \includegraphics[width=1.5\textwidth]{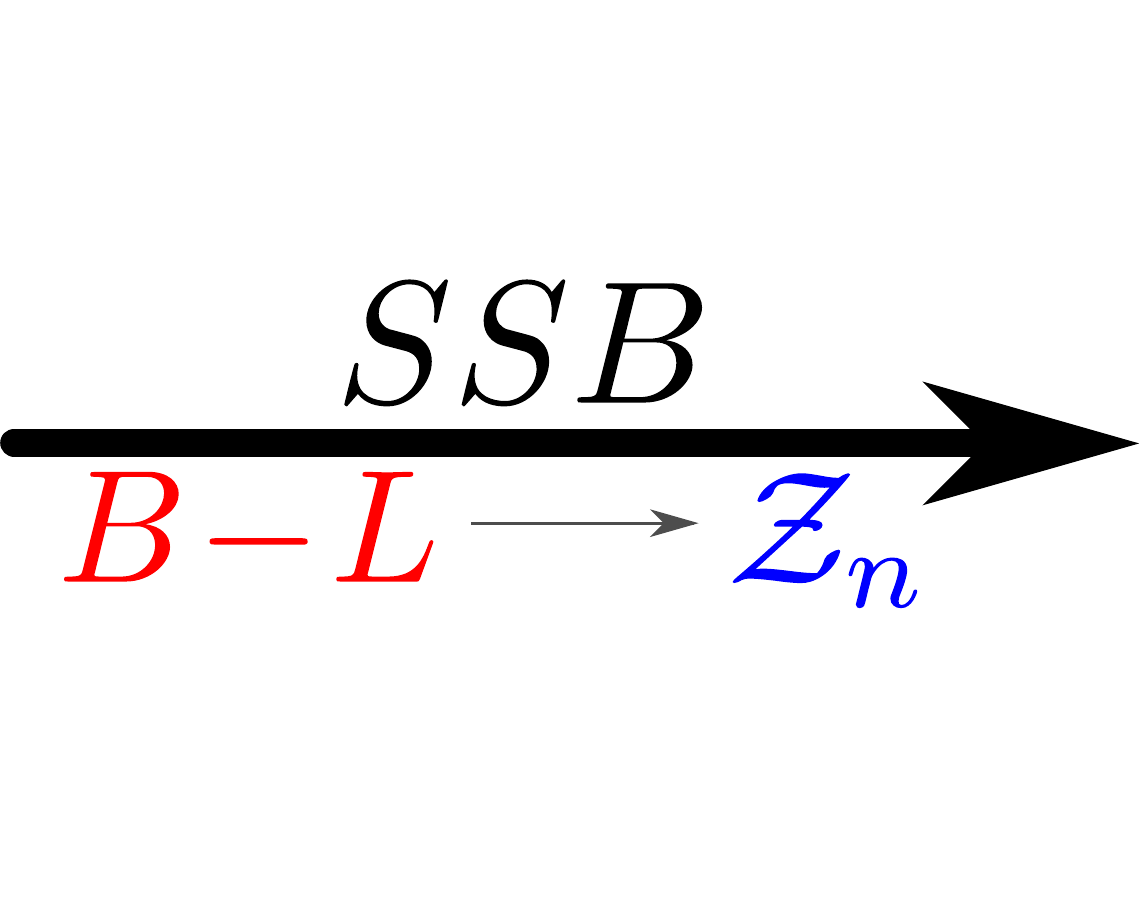}      \end{subfigure}
    ~ 
   \begin{subfigure}[b]{0.43\textwidth}
        \includegraphics[width=\textwidth]{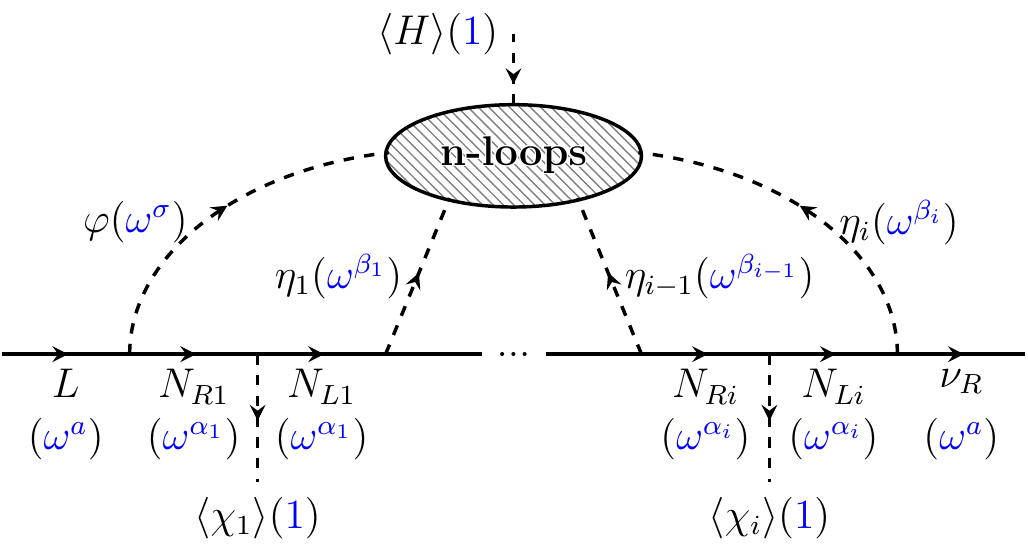}
        \caption{General residual $\mathcal{Z}_n$ charge assignment.}
        \label{fig:genZn}
    \end{subfigure} \\
    \caption{General charge assignment for any topology and its spontaneous symmetry breaking pattern.}\label{fig:gencase}
\end{figure}

\end{widetext}

Our general strategy is to use the chiral anomaly free solutions of $U(1)_{B-L}$ symmetry to generate loop masses for Dirac neutrinos and also have a stable dark matter particle mediating the aforementioned loop. Then, after symmetry breaking, once all of the scalars get a vacuum expectation value (vev), the $U(1)_{B-L}$ symmetry will be broken down to one of its $\mathcal{Z}_n$ subgroups, such that the dark matter stability and Dirac nature of neutrinos remain protected. This scheme is shown diagrammatically  in Fig.~\ref{fig:gencase}. 

In Fig.~\ref{fig:gencase} the SM singlet fermions $N_{Li}, N_{Ri}$, as well as the right-handed neutrinos $\nu_{R}$, have non-trivial chiral charges under $U(1)_{B-L}$ symmetry.\footnote{It is not necessary that all fermions $N_{Li}, N_{Ri}$ be chiral under $U(1)_{B-L}$ symmetry.} In order to generate the masses of these chiral fermions we have also added SM singlet scalars $\chi_i$ which also carry $U(1)_{B-L}$ charges. To complete the neutrino mass generation loop, additional scalars $\varphi, \eta_i$ are required. After spontaneous symmetry breaking (SSB) of the $U(1)_{B-L}$ symmetry, all of the scalars $\chi_i$ will acquire vevs that break the $U(1)_{B-L} \to \mathcal{Z}_n$ residual symmetry. The fermions $N_{Li}, N_{Ri}$ get masses through the vevs of the scalars $\chi_i$, while the neutrinos acquire a naturally small $n$-loop mass as shown in Fig. \ref{fig:gencase}.

In order to satisfy all of the requirements listed above, several conditions must be applied. First of all, the model should be anomaly-free:
\begin{itemize}[label=\textcolor{orange}{\textbullet}]
\item The chiral charges of the fermions must be taken in such a way that the anomalies are canceled.
\end{itemize}
In order to obtain non-zero but naturally small Dirac neutrino masses we impose the following conditions:
\begin{itemize}[label=\textcolor{blue}{\textbullet}]
\item The tree-level Yukawa coupling $\bar{L} \tilde{H} \nu_R$ should be forbidden. This implies that apart from the SM lepton doublets $L_i$ no other fermion can have $U(1)_{B-L}$ charge of $\pm 1$. Furthermore, to ensure that the desired loop diagram gives the dominant contribution to the neutrino masses, all lower loop diagrams should also be forbidden by an appropriate choice of the charges of the fields. 
 \item The operator leading to neutrino mass generation, i.e. $\bar{L} H^c \chi_1 \dots \chi_i \nu_{R}$, should be invariant under the SM gauge symmetries as well as under $U(1)_{B-L}$. Following the charge convention of Fig.~\ref{fig:gencase}, the charges of the vev carrying scalars $\chi_i$ should be such that $\sum_i \zeta_i = -1 - \ell$ .
 \item All of the fermions and scalars running in the neutrino mass loop must be massive. Since the fermions will be in general chiral, this mass can only be generated via the coupling with a vev carrying scalar. For example, in the diagram in Fig.~\ref{fig:gencase} we should have $-x_i + x'_i + \zeta_i = 0$.
\item To protect the Diracness of neutrinos, all of the Majorana mass terms for the neutrino fields at all loops must be forbidden in accordance with Eq.~\eqref{evenzndir}.
\end{itemize}

Additionally, for dark matter stability, we impose the following conditions:

\begin{itemize}
 \item After SSB, the $U(1)_{B-L}$ symmetry is broken down to a $\mathcal{Z}_n$ subgroup. Only even $\mathcal{Z}_n$ subgroups with $n > 2$ can protect dark matter stability. The odd $\mathcal{Z}_n$ subgroups invariably lead to dark matter decay.\footnote{For odd $\mathcal{Z}_n$ subgroups, there will always be an effective dark matter decay operator allowed by the residual odd $\mathcal{Z}_n$ symmetry. Even then it is possible that such an operator cannot be closed within a particular model, thus pinpointing the existence of an accidental symmetry that stabilizes dark matter. Another possibility is that the dark matter candidate decays at a sufficiently slow rate. Thus for residual odd $\mathcal{Z}_n$ symmetries, one can still have either a stable dark matter stabilized by an accidental symmetry or a phenomenologically viable decaying dark matter. In this paper, we will not explore such possibilities.} The symmetry breaking pattern can be extracted as follows. First all the $U(1)$ charges must be rescaled in such a way that all the charges are integers and the least common multiple (lcm) of all of the rescaled charges is $1$. Defining $n$ as the least common multiple of the charges of the scalars $\chi_i$, it is easy to see that the $U(1)$ will break to a residual $\mathcal{Z}_n$. This $n$ must be taken to be even as explained before, i.e. $n \equiv $ lcm$(\zeta_i) \in 2\mathbb{Z}$. 
 \item Dark sector particles should neither mix with nor decay to SM particles or to vev carrying scalars.
\item There are two viable dark matter scenarios depending on the transformation of the SM fermions under the residual symmetry. 
\begin{itemize}
\item When all SM fields transform as even powers of $\omega$, where $\omega^n = 1$, under the residual $\mathcal{Z}_n$, the lightest particle transforming as an odd power will be automatically stable, irrespective of its fermionic or scalar nature. We will show an explicit example of this simple yet powerful idea later.
\item In the case in which all SM fermions transform as odd powers of the residual subgroup, it can be shown that all of the odd scalars and the even fermions will be stable due to a combination of the residual $\mathcal{Z}_n$ and Lorentz symmetry. 
\end{itemize}
\end{itemize}

Given the long list of requirements, most of the possible solutions that lead to anomaly cancellation fail to satisfy some or most of them. 
Still we have found some simple one-loop and several two-loop solutions that can satisfy all of the conditions listed above.

In this paper, we demonstrate the idea for a simple solution in which the $U(1)_{B-L}$ symmetry is broken down to a residual $\mathcal{Z}_6$ symmetry. 
However, in general, many other examples with different residual even $\mathcal{Z}_n$ symmetries can be found by applying the given framework.

\begin{center}{\bf Realistic example}\end{center}

\begin{widetext}

 \begin{figure}[th]
    \centering
    \begin{subfigure}[b]{0.4\textwidth}
        \includegraphics[width=\textwidth]{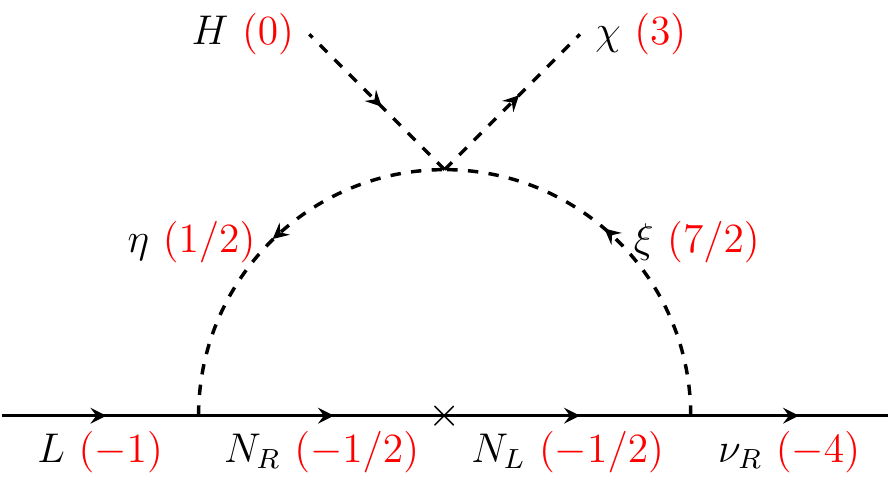}
        \caption{$U(1)_{B-L}$ charge assignment.}
        \label{fig:gull}
    \end{subfigure}
    ~ 
    \begin{subfigure}[t]{0.1\textwidth}
    \vspace*{-3.5cm}
    \hspace*{-0.4cm}
   \includegraphics[width=1.4\textwidth]{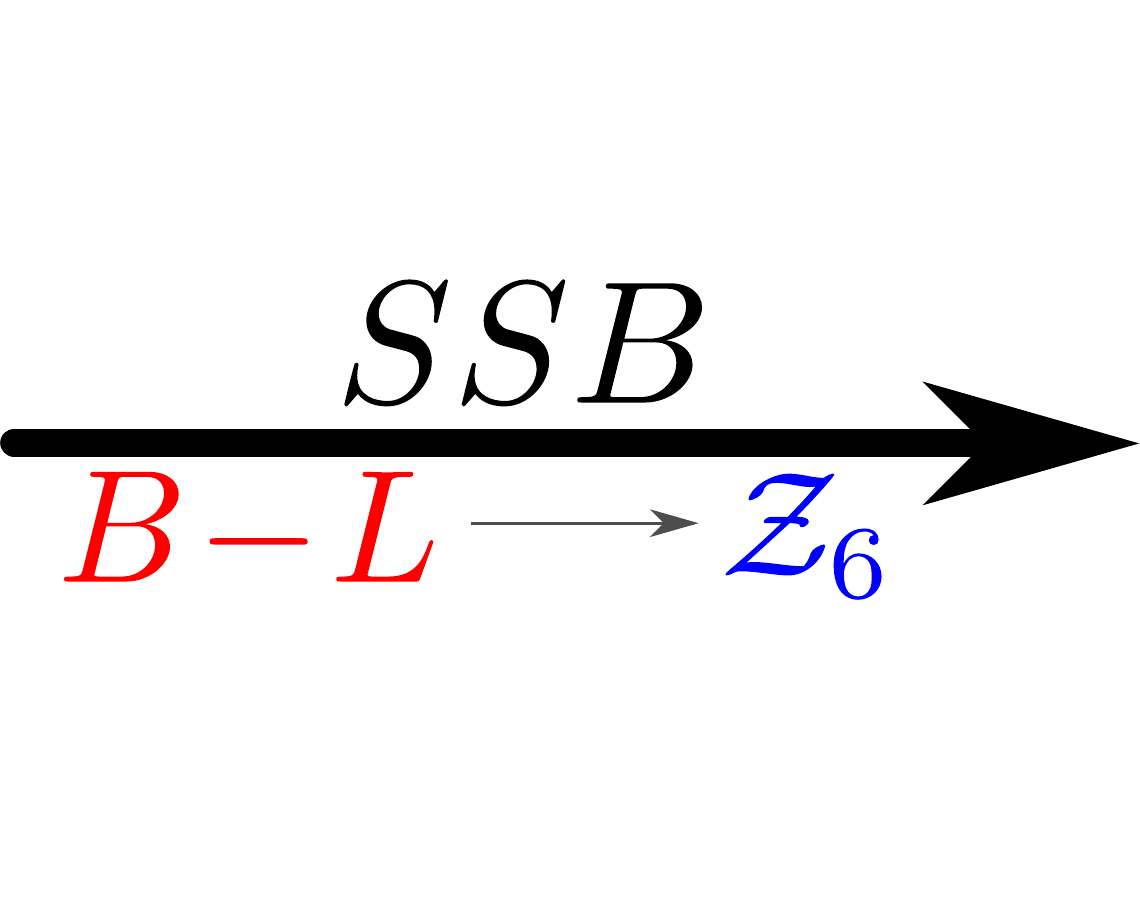}      \end{subfigure}
    ~ 
   \begin{subfigure}[b]{0.4\textwidth}
        \includegraphics[width=\textwidth]{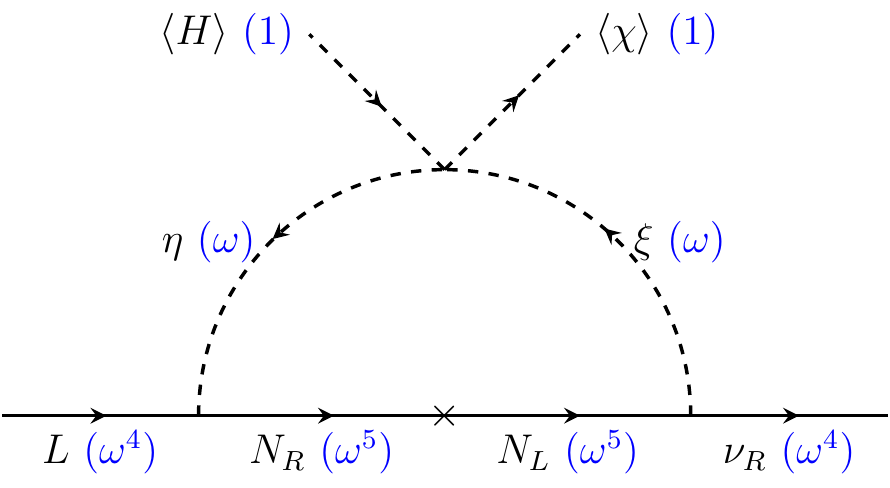}
        \caption{Residual $\mathcal{Z}_6$ charge assignment.}
        \label{fig:mouse}
    \end{subfigure} \\
    \caption{Charge assignment for the example model and its spontaneous symmetry breaking pattern.}\label{fig:z6}
\end{figure}
\end{widetext}

Let us consider an extension of the SM by adding an extra Higgs singlet $\chi$ with a $U(1)_{B-L}$ charge of $3$, along with an scalar doublet $\eta$, a singlet $\xi$ and two vector-like fermions $N_{L_l}$ and $N_{R_l}$, with $l=1,2$, all carrying non-trivial $U(1)_{B-L}$ charges as shown in Table \ref{modelZ6} and depicted in Fig.~\ref{fig:gull}.
\begin{table}
\begin{center}
\begin{tabular}{| c || c | c | c || c |}
  \hline 
&   Fields            &    $SU(2)_L \otimes U(1)_Y$            &     $U(1)_{B-L}$                       & 
  $\mathcal{Z}_{6}$                              \\
\hline \hline
\multirow{4}{*}{ \begin{turn}{90} Fermions \end{turn} } &
 $L_i$        	  &    ($\mathbf{2}, {-1/2}$)       &   {\color{red}${-1}$}    	  &	 {\color{blue}$\omega^4$}                     \\	
&   $\nu_{R_i}$       &   ($\mathbf{1}, {0}$)      & {\color{red} $({-4},{-4},\,{5})$ }   &  	 {\color{blue}($\omega^4, \omega^4, \omega^4)$}\\
&   $N_{L_l}$    	  &   ($\mathbf{1}, {0}$)      & {\color{red}${-1/2}$ }   &    {\color{blue} $\omega^5$}     \\
&  $N_{R_l}$     	  &  ($\mathbf{1}, {0}$) 	     & {\color{red} ${-1/2}$ } &  {\color{blue}$\omega^5$}     \\
\hline \hline
\multirow{4}{*}{ \begin{turn}{90} Scalars \end{turn} } &
 $H$  		 &  ($\mathbf{2}, {1/2}$)      &  {\color{red}${0}$ }    & {\color{blue} $1$}    \\
& $\chi$          	 &  ($\mathbf{1}, {0}$)        &  {\color{red}${3}$ }  &  {\color{blue} $1$}     \\		
& $\eta$          	 &  ($\mathbf{2}, {1/2}$)      &  {\color{red}${1/2}$}    &  {\color{blue}$\omega$}       \\
& $\xi$             &  ($\mathbf{1}, {0}$)        &  {\color{red}${7/2}$}      &	{\color{blue}$\omega$} \\	
    \hline
  \end{tabular}
\end{center}
\caption{Charge assignment for all of the fields. $\mathcal{Z}_6$ is the residual symmetry in this example, with $\omega^6=1$.}
 \label{modelZ6} 
\end{table}

The neutrino interactions are described by the following Lagrangian, 
\begin{equation}
{\cal L}_\nu=y_{il}\bar{L}_i\tilde{\eta}N_{R_l}+y'_{li}\bar{N}_{L_l}\nu_{R_i}\xi+M_{lm} \bar{N}_{R_l}N_{L_m}+h.c.,
\label{eq:YukInt}
\end{equation}
where $\tilde{\eta} = i \tau_2 \eta^*$, with the indices $i = 1,2,3$ and $l,m = 1,2$. The relevant part of the scalar potential for generating the Dirac neutrino mass is given by
\begin{equation}
\label{Veps}
{\cal V}\supset m_{\eta}^2 \eta^\dagger \eta + m_\xi^2 \xi^\dagger \xi + (\lambda_D H^\dagger \eta \chi \xi^* + h.c.),
\end{equation}
where $\lambda_D$ is a dimensionless quartic coupling. 
 
After spontaneous symmetry breaking of $U(1)_{B-L}$, the scalar $\chi$ gets a vev $\vev{\chi}=\text{u}$, giving mass to two neutrinos through the loop depicted in Fig.~\ref{fig:z6}. Note that only $\nu_{R_1}$ and $\nu_{R_2}$ can participate in this mass generation due to the chiral charges $(-4, -4, \, 5)$, i.e. $y'_{l3}=0$ in Eq.~\eqref{eq:YukInt}. The third right-handed neutrino $\nu_{R_3}$ remains massless and decouples from the rest of the model, although it is trivial to extend this simple model to generate its mass. 

The neutral component of the gauge doublet $\eta$ and the singlet $\xi$ are rotated into the mass eigenbasis with eigenvalues $m^2_i$ in the basis of ($\xi$,\,$\eta^0$). The neutrino mass matrix is then given in terms of the one-loop Passarino-Veltman function $B_0$ \cite{Passarino:1978jh} by,
 \begin{equation}
    (M_\nu)_{\alpha\beta} \sim \frac{1}{16\pi^2} \frac{\lambda_D \text{v} \text{u}}{m^2_\xi-m^2_\eta} y_{\alpha k} y'_{k\beta}  M_k \sum\limits_{i=1}^2 (-1)^i B_0(0,m_i^2,M_k^2),
    \label{numass}
\end{equation}

where $M_k$ ($k=1,2$) are the masses of the Dirac fermions $N_k$ and $\vev{H}=\text{v}$ the Standard Model vev.

As a benchmark point, we can take the internal fermion to be heavier than the scalars running in the loop, one of which will be the dark matter candidate. Then, Eq.~\eqref{numass} can be approximated by,
\begin{equation}
    m_\nu \sim \frac{1}{16 \pi^2} \frac{v u}{M} y y' \lambda_D.
\end{equation}
For comparison, we can take the Yukawa couplings to be of order $10^{-2}$ and the quartic coupling $\lambda_D \sim 10^{-4}$, like in the original scotogenic model \cite{Ma:2006km}. We can also take neutrino masses to be of order $0.1$ eV and $u \sim v$. With these choices, we can find the mass scale of the neutral fermions,
\begin{equation}
M \sim \frac{1}{16 \pi^2} \frac{v u}{m_\nu} y y' \lambda_D \sim 10^{4} \text{GeV}.
\end{equation}

Compared with the type-I seesaw scale $M \approx y^2 \frac{v^2}{m_\nu} \sim 10^{10}$ GeV we can see a five order of magnitude suppression coming from the loop and the possibility of a broader parameter space.

It is worth mentioning that since the $U(1)_{B-L}$ is anomaly free, it can be gauged. Then the physical Nambu-Goldstone boson associated to the dynamical generation of the Dirac neutrino mass \cite{Bonilla:2016zef} is absent.
   
\begin{figure}[h!]
\includegraphics[width=0.33\textwidth]{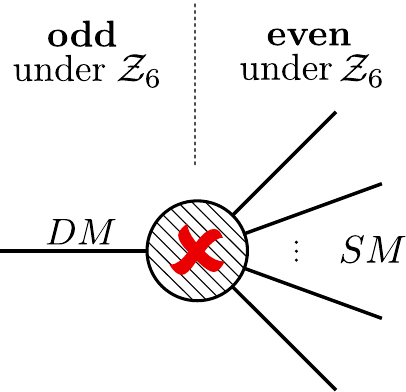}
\caption{The decay of dark matter (odd under $\mathcal{Z}_6$) to SM particles (all even under $\mathcal{Z}_6$) is forbidden by the residual $\mathcal{Z}_6$ symmetry. This argument can be generalized to any even $\mathcal{Z}_n$ symmetry.}
\label{U1mod}
\end{figure}

Regarding dark matter stability in this particular model, we can see that the lightest particle inside the loop is stable. This is true for both the fermionic and scalar dark matter candidates. As can be seen in Fig.~\ref{fig:mouse}, all of the internal loop particles are odd under the remnant $\mathcal{Z}_6$, while all of the SM particles are even. Therefore any combination of SM fields will be even under the residual subgroup, forbidding all effective operators leading to dark matter decay as shown graphically in Fig.~\ref{U1mod}.
   
To summarize, we have shown that by using the $U(1)_{B-L}$ symmetry already present in the Standard Model, it is possible to address the dark matter stability and relate it to the smallness of Dirac neutrino masses.
We have described a general framework in which these features are realized by exploiting the anomaly free chiral solutions of a global $U(1)_{B-L}$. This framework can be utilized in a wide variety of scenarios. 
We have presented a particular simple realization of this idea where neutrino masses are generated at the one-loop level and the $U(1)_{B-L}$ symmetry is broken spontaneously to a residual $\mathcal{Z}_6$ symmetry. 
The framework can also be used in models with higher-order loops as well as in cases where $U(1)_{B-L}$ symmetry is broken to other even $\mathcal{Z}_n$ subgroups.
Since the $U(1)_{B-L}$ is anomaly free, it can be gauged in a straightforward way, giving a richer phenomenology from the dark matter and collider point of view.

\begin{acknowledgments}

E.P. would like to thank the groups AHEP (IFIC) and TUM for their hospitality during his visits. R.S. would like to thank IFUNAM for the warm hospitality during his visit. This work is supported by the Spanish grants FPA2017-85216-P (AEI/FEDER, UE), Red Consolider MultiDark FPA2017-90566-REDC and PROMETEO/2018/165 (Generalitat  Valenciana). The work of C.B. was supported by the Collaborative Research Center SFB1258. S.C.C. is supported by the Spanish grant BES-2016-076643. R.C. is supported by the Spanish grant FPU15/03158. E.P. is supported in part by DGAPA-PAPIIT IN107118, CONACYT CB2017-2018/A1-S-13051 (Mexico), the German-Mexican research collaboration grant SP 778/4-1 (DFG) and 278017 (CONACyT) and PIIF UNAM.

R.S. will like to dedicate this paper to his grandparents, who despite not knowing what he is doing, never stopped believing in him.

\end{acknowledgments}



%

\end{document}